# Ingroup bias is prevalent in user reports of hate and abuse online

Florence E. Enock[1,4], Helen Z. Margetts[1,2] and Jonathan Bright[3]

[1] Oxford Internet Institute, Stephen A. Schwarzman Centre for the Humanities, University of Oxford, Radcliffe Observatory Quarter, Oxford. OX2 6GG.

[2] Data Science Institute, The London School of Economics and Political Science, Houghton Street, London. WC2A 2AE.

[3] Pattrn.ai, 264 Banbury Road, Oxford. OX2 7DY.

[4] Public Policy Programme, The Alan Turing Institute, The British Library, 96 Euston Road, London. NW1 2DB.

**Author contact details**

*Florence Enock: florence.enock@oii.ox.ac.uk (Orcid: 0000-0003-2108-4891)

Helen Margetts (Orcid: 0000-0003-4597-8283)

Jonathan Bright (Orcid: 0000-0002-1248-9275)

*Corresponding author

**Acknowledgements**

This work was supported by a British Academy/Leverhulme Small Research Grant awarded to FE (reference: SRG2223\230966) and core funding from The Alan Turing Institute.

Accepted for publication in *Royal Society Open Science* (June 2026). This is the accepted manuscript prior to copyediting and typesetting.

## Abstract

The prevalence of online hate and abuse is a pressing global problem. While tackling such societal harms is a priority for research across the social sciences, it is a difficult task, in part because of the magnitude of the problem. People's engagement with reporting mechanisms ('flagging') online is an increasingly important part of monitoring and addressing harmful content at scale. However, users may not flag content routinely enough, and when users do engage, they may be biased by group identity and political beliefs. Across five well-powered and pre-registered online experiments, we examine the extent of ingroup bias in people's flagging of hate and abuse in four different intergroup contexts: political affiliation, vaccination opinions, beliefs about climate change, and stance on abortion rights. Overall, participants reported abuse reliably, with approximately half of the abusive comments in each study reported. However, a pervasive ingroup bias was present whereby across studies, participants were between 17% and 63% more likely to flag abuse directed at the ingroup than at the outgroup. Our findings offer new insights into the nature of user flagging online, an understanding of which is crucial for enhancing user intervention against online hate and thus ensuring a safer online environment.

## Main text

The prevalence of online hate and abuse continues to be a pressing social issue. While definitions vary (1), online abuse generally refers to content containing threats, incitements to violence, dehumanizing slurs or derogatory attacks, while online hate typically refers to content of this sort which targets specific and protected identities, such as ethnicity, religion or gender (1–4). While it is difficult to accurately measure how much hate and abuse there is online, there is evidence from multiple sources to suggest the problem is widespread (5). A recent UK survey found that 60% of UK adults have seen hateful content online before (6) and in the US, survey research found that 41% of adults in America had directly experienced online harassment, with a quarter reporting personal experience with stalking, harassment, and physical threats (7). Platform reports also highlight the scale of the problem – for example, in the first quarter of 2025, Meta reports taking action over 3.4 million pieces of content containing hate (though the prevalence reported has dropped substantially since 2021) (8). High levels of online hate and abuse are problematic for many reasons – exposure can cause severe harm to the psychological wellbeing of targets, with experiences linked to depression, anxiety, fear, low self-esteem and escalation of self-harm (9,10) and online hate can also provoke and justify violent attacks offline (11–13). A recent systematic review found that exposure to online hate can worsen attitudes and stereotypes towards targeted groups, reduce intergroup trust, fuel the spread of further hateful content, and damage psychological wellbeing and life satisfaction among those exposed (3). As such, working to tackle such content is a priority area for research across the social sciences, but it is a difficult task, in part because of the magnitude of the problem.

Many interventions have been developed in response to the problem of online hate and abuse. These span those directly implemented by platforms, such as automated content moderation, algorithmic downranking of content and deplatforming perpetrators (14–16), to initiatives designed to raise awareness amongst users, such as media literacy courses and public awareness campaigns (12). Community-based approaches often emphasise the importance of bystander behaviour, encouraging users to take an active role through safely challenging harmful content. Counterspeech, in which users directly challenge harmful content by expressing disagreement, correcting misinformation or supporting targeted groups, is one such bystander intervention that has gained research attention in recent years (17). Experimental and observational work suggests that counterspeech may be effective at mitigating the negative effects of online hate under certain conditions, for example when it is delivered with empathy, supported by social norms, or when it comes from ingroup members (18,19). However, counterspeech is not consistently effective and may sometimes even backfire, escalating or amplifying the visibility of hate (20–22). From a user perspective, actively producing counterspeech requires time, emotional labour, and a willingness to expose oneself to potential retaliation, potentially making many users unwilling or unable to intervene in this way.

Reporting (or 'flagging') content is a comparatively low cost form of bystander intervention – it allows users to signal the presence of harmful content or violation of community standards directly to platforms quickly, confidentially, and without fear of repercussion from perpetrators. There is widespread agreement that user engagement with these reporting mechanisms (known as 'flagging') has the potential to play a crucial role in allowing platforms to address harmful language online at scale. However, while all mainstream platforms typically allow users to flag content which they believe to be inappropriate, offensive, or as breaching community guidelines, there remain concerns that flagging mechanisms do not currently work as well as is needed (23). There are two key factors which may underlie the efficacy of user flagging: the extent of user engagement (how much people report harmful content), and the quality or accuracy of such flags (how biased flags are by social and intergroup contexts).

Relatively little is known about how much content overall is flagged by users online because platforms do not typically make this information available. Survey data suggests that awareness of such functions is high, with 86% of social media users aware of the ability to report content (24). The same survey suggests that use is promising, with 60% of those who were aware of being able to report choosing to do so at least once. However, data elsewhere suggests user engagement with reporting mechanisms to be lower – work from Ofcom shows in 2023, only 20% of internet users in the UK had reported the last piece of harmful content they came across (25). Together, these findings suggest that while a majority may have flagged content at some point, few do so regularly and in response to all harms.

In terms of the quality or accuracy of flags, there is preliminary evidence to suggest that when users do engage, flagging can be biased by group identity and political beliefs. Some work shows that when it comes to reporting content as misinformation, people are more likely to report statements opposing their own political views as false (even when they are factually accurate) (26). These findings suggest that flagging can be driven as much by intergroup bias as by concern about content accuracy or safety. This account dovetails with older experimental results from an offline context, showing that when US-based participants were shown a video of a protest and told that it depicted liberals opposing military recruitment, Republicans were more in favour of police intervention than Democrats, while the opposite was true when participants were told video showed a conservative protest (opposing an abortion clinic). Faced with identical visual information, people's endorsement of intervention rested on the ideological congruence of the content (27,28). A body of work in social cognition similarly shows that partisanship and intergroup biases shape multiple levels of cognition and types of behaviour (29–33). It is possible, then, that user engagement with flagging hate and abuse online may be similarly biased by social identity. However, no work to date has systematically examined how such biases impact user engagement with reporting hate and abuse online.

Given the importance of user engagement for tackling harmful language online at scale, it is crucial to assess how platforms can improve their reporting mechanisms, both in encouraging users to flag harmful content more routinely, and in making sure the right kind of content is flagged. This is particularly important because if flagging behaviours are strongly influenced by group ideologies and social biases, then the usefulness of user-initiated flags in tackling online hate will be undermined. It is only beneficial to work to encourage uptake if people generally flag the right kind of content. If ingroup biases are overly influential in such behaviours, interventions should first work to enhance the quality and accuracy of user flags.

Here, we examine the extent to which user flagging is a reliable signal for detecting hate and abuse. Using mock online environments, we measure how much content people flag overall, along with the extent to which ingroup biases influence flagging behaviours. Across five studies, we examine patterns of flagging of hate and abuse in four different intergroup contexts in the US: political affiliation, vaccination opinions, beliefs about climate change, and stance on abortion rights. In each study, participants are presented with a series of mock social media comments and asked to flag any which they believe to be offensive and would typically wish to report if they came across them in daily life. The comments vary across language type (abuse or critical) and target (ingroup or outgroup). In all studies, participants are recruited to equally represent both sides of each issue, for example, in Study 1, half identify as Democrats and half as Republicans. In Studies 1-4a, participants each see comments directed at one of the two group targets (e.g., in Study 1, either Democrats or Republicans depending on which between-subjects condition participants fall into). In Study 4b, participants see comments directed at both groups in the same 'feed'. If flagging is driven primarily by language severity (a rational, non-biased account), we would expect a main effect of language type only, with abuse receiving more flags than critical speech regardless of the target. If flagging behaviours are biased by social identity, we expect to observe an effect of target group such that participants are more likely to flag content directed at the ingroup than the outgroup. By including balanced samples and content directed at both sides of each issue, we are also able to explore nuances in flagging patterns across several different groups and domains.

**Data collection and open science**

All data collection took place online and studies were created and administered using Qualtrics (https://www.qualtrics.com). Data for Experiments 1-3 were collected between August and October 2023, and data for Experiments 4a and 4b were collected in March 2023. Participants were recruited through Prolific (https://www.prolific.com) and different participants took part in each study. Informed consent was obtained at the start of each session according to approved ethical procedures. Participants were compensated at an approximate rate of £9 per hour. Links to pre-registration documents, additional study materials, raw data and R analysis scripts for each study can be found at the OSF page for this work: https://osf.io/tx9cs.

**Study 1: Ingroup bias in flagging politically motivated abuse**

In our first experiment, we aim to understand the extent to which people typically flag online abuse in the context of politically motivated discussion in the US, along with the extent of ingroup bias within such flagging behaviours. We test the extent to which people flag abuse compared to critical speech, and whether people are more likely to flag when comments are directed at the political ingroup compared to the political outgroup. We choose a US-based political context because a body of work suggests that online debates around this topic can be divisive and heavily polarising (34,35).

**Methods**

**Participants**

For the pre-registered 2*2*2 mixed ANOVA, a power analysis using G*power 3.1 found a minimum N of 316 to be necessary to detect effects of interest with a small to medium effect size ($\eta_p^2 = .04$), alpha of .05 and power of .95.

340 participants completed the study. Participants were all 18 or over, fluent in English, and based in the US. In line with our pre-registration, we excluded 24 participants that failed at least one of the two attention checks. Of the final sample of 316, 156 participants identified as female, 154 as male, 5 as non-binary and 1 self-described. Participants were aged from 18 to 78 with a mean age of 40.7 (SD = 13.9). 161 participants identified as Democrats, and 155 participants identified as Republicans.

**Flagging task**

To measure the extent of ingroup bias in flagging behaviours, we developed a novel paradigm in which participants read through a series of online posts designed to look like user-generated comments below a news story. Of 48 experimental items, 24 constituted abuse and 24 constituted mild criticism. The abusive comments were adapted from a real dataset tracking harmful language online (36) and they were designed to include a various types of harmful language, including incitements of violence, dehumanizing slurs, and general abuse (11).

The target of the comments was either the Democrats or the Republicans depending on which between-subjects condition participants were in. Examples of abusive comments were, ‘Hang all [Republican/Democrat] supporters’ and ‘The [Republicans/Democrats] are truly vermin in human form’. Examples of critical comments were, ‘[Republicans/Democrats] focus on the wrong issues’ and ‘I don’t really trust what the [Republicans/Democrats] say’ (see Table 1). There were also five filler items which were neutral in tone and did not reference either political party (e.g., ‘I am optimistic about what the future holds’), along with two attention checks.

Below each comment was a flag icon, similar in design to those which may be found on social media platforms. Participants were asked to use the flag icon to report any comments which they found offensive and would typically wish to flag if they came across them in everyday life. If participants clicked the flag icon, it became highlighted, meaning they had chosen to 'report' the content. If participants clicked the item in error, they could click it again to unflag. All experimental and filler items were presented in a random order for each participant and the attention checks were placed approximately a third and two thirds of the way through. The flagging task was presented in Qualtrics, with posts presented sequentially on a single scrolling page and with each post shown in a consistent layout with the flag option underneath the text for each. While this environment does not reproduce all features of a typical platform, it provides a controlled experimental setting for testing user reactions to online content. An example screenshot of the Qualtrics interface from Experiment 1 is provided in Figure S1 in the Supplementary Information.

To set the context for the flagging task, participants first read a short excerpt from a news story discussing the positioning of Democratic and Republican candidates for the upcoming election in the US in 2024. The news article was chosen to be as neutral as possible and to mention both parties to an equal degree (examples of these for all studies are found on the OSF page for this work). Data collection took place in September 2023, during a period when a Democratic president was in office and more than a year remained until the next U.S. presidential election.

**Table 1:** Mock social media posts used in the flagging task. Each of the comments could describe either of the political parties, depending on which condition participants were in. Half of the participants saw comments directed at ingroup members and half saw comments directed at outgroup members, depending on which political ideology participants identified with. The abusive comments were adapted from real examples of online abuse and sought to include different types of harmful language (incitements of violence and dehumanizing slurs). Comments were presented in a randomised order for each participant.

| Item | Abusive | Critical |
|---|---|---|
| 1 | [Target] voters should be abused | The [Target] aren't doing enough for our children |
| 2 | Hang all [Target] supporters | The [Target] don't really care about equality |
| 3 | The [Target] should be gassed | [Target] focus on the wrong issues |
| 4 | Calling for attacks on all [Target] c*nts - who can get one put in hospital first | [Target] need to tone down the wording of their campaigns |
| 5 | [Target] voters are detestable rats from hell | I don't trust what [Target] say |
| 6 | [Target] are nothing but a lying sack of scum | [Target] are not capable |
| 7 | The [Target] lefties are truly vermin in human form | [Target] voters can be pretty annoying |
| 8 | [Target] supprters you are disgusting, vile, immoral cockroaches | [Target] come across as so hypocritical sometimes |
| 9 | [Target] are lying, cheating criminals that will burn in hell | [Target] voters are not very intelligent |
| 10 | [Target] voters are arrogant hypocritical c*nts | [Target] voters don't understand economics |
| 11 | Anyone supporting [Target] is a f*cking duplicitous evil tw*t | [Target] don't reflect American values |
| 12 | [Target] voters you are spineless pieces of sh*t | [Target] supporters are boring |
| 13 | Round up and kill anyone that votes [Target] | People who vote [Target] are harming our country |
| 14 | Personally, I think we should just burn the [Target], who's with me? | The [Target] just don't take enough action |
| 15 | grievous bodily harm is allowed if it's against a [Target], right? | The [Target] are simply weak |
| 16 | [Target] lefties are fair game for a battering in my opinion | [Target] supporters aren't living in the real world |
| 17 | [Target] you are the scum of our society, pure cockroaches | People voting [Target] don't really know what they are doing |
| 18 | [Target] = corrupt f*cking worms | [Target] policies don't even make sense |
| 19 | Advocates of [Target]ic policies are not humans they are a subpar species | The [Target] are too forceful in their messages |
| 20 | The [Target] are a a vicious plague in our society | People supporting the [Target] are misguided |
| 21 | [Target] campaigners = ignorant useless pricks | [Target] campaigners don't care about our safety |
| 22 | [Target] you are nothing but theiving, lying criminals | [Target] supporters lack integrity in my experience |
| 23 | [Target] - you are a disgrace to the human race | [Target] voters just aren't educated properly on the right matters |
| 24 | Anyone supporting the [Target] is a f*cking disgrace - pathetic ignorant bast*ards | I find [Target] voters a bit self-righteous |

## Procedure

Participants were informed that the study was designed to help us understand the ways in which people report content which may be harmful online. After informed consent was taken, participants answered demographic questions about age, gender, ethnicity, and political orientation (measured using a continuous slider from left to right). Screening for eligibility was through Prolific, but participants were also asked about their US political affiliation (Democrat or Republican) to ensure that their response matched their affiliation recorded on Prolific and that they were in the correct version of the experiment. If they indicated 'Democrat' and were in the Republican version of the study, or vice versa, they were screened out on the basis that the response was inconsistent with their Prolific pre-screening.

Following demographics and pre-screening, participants were presented with instructions for the main study. They were reminded that many online platforms allow people to report content which violates community standards, such as posts that include hate or violence. They then read:

> *You will now read a news article about the next presidential election. Following this, you will see several user-generated comments. Some of these comments may be offensive and include hateful, abusive or threatening content. Not all comments include language of this type. As you read through the comments, please use the flag icon to report any comments which you find offensive and would typically wish to flag if you came across these online in daily life.*

Participants were shown the flag icon, and told that if they click the icon, it becomes highlighted, meaning they have chosen to report the content. After the instructions, participants read the news article to set the context for the comments. After reading the news article, participants were reminded of the task instruction and then read through the series of mock user-generated comments.

After completing the flagging task, participants had the opportunity to provide feedback using a free-text box and were taken to the debrief. In the debrief, participants were provided with additional information about the purpose of the study and were also informed that the comments they read were not real but had been based on typical examples of hate and abuse. Participants were then redirected back to Prolific for payment. Most participants took approximately seven minutes to complete the study.

**Design and data analysis**

The type of comments participants saw took a two-by-two structure: all participants saw example of both abuse and mild criticism (Language type), and half of the participants saw comments directed at Republicans, while the other half saw comments directed at Democrats (Target). As half of participants identified as Republicans and half as Democrats, the Target was an ingroup or an outgroup depending on participants' own political affiliation. This results in a 2 (Language type: Abuse/Criticism, within-subjects)*2 (Target: Ingroup/Outgroup, between-subjects)*2 (Political affiliation: Republican/Democrat, between-subjects) mixed factorial design. As pre-registered, we initially calculated the proportion of flags made in each experimental condition and conducted mixed ANOVAs on these proportions to test for differences in flagging between the experimental conditions. We report these as percentage flagged in our results summaries. All ANOVAs reported here were conducted using Type III sums of squares.

## Results and discussion

In line with our pre-registered analysis plan, a 2(Language type: Abuse/Critical, within subjects) x 2(Target: Ingroup/Outgroup, between subjects) x 2 (Political affiliation: Democrat/Republican, between subjects) mixed ANOVA measured differences between proportion of comments flagged in each condition.

There was a main effect of language type, $F(1, 312) = 1020.46$, $p < .001$, $\eta_p^2 = .766$, showing abuse was flagged to a significantly greater extent than criticism (47.1% of overall abuse was flagged compared to 1.3% of criticism). In line with our prediction, there was also a main effect of target, $F(1, 312) = 6.58$, $p = .011$, $\eta_p^2 = .021$, showing flagging rates were higher for comments directed at the ingroup than at the outgroup (50.5% of ingroup abuse was flagged compared to 43.8% of outgroup abuse, and 2% of ingroup criticism was flagged compared to 0.5% of outgroup criticism). There was no significant main effect of political affiliation overall, $F(1, 312) = 1.48$, $p = .224$, $\eta_p^2 = .005$, showing Democrats and Republicans flagged to a similar extent.

There was a significant two-way interaction between political affiliation and target, $F(1, 312) = 6.45$, $p = .012$, $\eta_p^2 = .020$, suggesting that the effect of target group was dependent on political affiliation. Pairwise comparisons showed that Republicans flagged abuse directed at the ingroup to a greater extent than abuse directed at the outgroup ($M_{IG} = 54.9\%$, $M_{OG} = 41.8\%$, $p = .002$), while Democrats flagged abuse directed at both groups to a similar extent ($M_{IG} = 46.1\%$, $M_{OG} = 45.6\%$, $p = .901$). Similarly, Republicans flagged criticism directed at the ingroup to a greater extent than criticism directed at the outgroup ($M_{IG} = 3.7\%$, $M_{OG} = 0.3\%$, $p = .013$), while Democrats flagged criticism directed at both groups to a similar extent ($M_{IG}$ 0.3%, $M_{OG}$ = 0.8%, $p$ = .453). No other interactions were significant (all $ps$ > .05).

Overall, results from the mixed ANOVA show that comments directed at the ingroup, both abusive and critical, were flagged to a greater extent than equivalent comments directed at the outgroup, but that this ingroup bias in flagging was only displayed by Republicans. However, it is important to note that the assumptions underlying the ANOVA were not fully met, likely owing to the proportional nature of the outcome variables. Specifically, residual diagnostics indicated departures from normality, and homogeneity of variance was violated for the critical language condition. To better account for the binary nature of trial-level flagging data and participant-level variability, we complemented our pre-registered ANOVA with generalized linear mixed models (GLMMs) using logistic regression.

**Additional analysis: Mixed-Effects Logistic Regression**

The GLMM included fixed effects of language type (abuse vs. criticism), target (ingroup vs. outgroup), and political affiliation (Democrat vs. Republican), along with all interactions, using sum coding for the categorical factors. A random intercept for participant was included to account for within-subject variance. We did not model within-subjects change in the dependent measure across trials because all items were presented in a random order for each participant and presentation order was not retained in the dataset.

Results from the GLMM supported our key findings: there was a significant main effect of language type, with abuse significantly more likely to be flagged than criticism (Odds ratio, OR = 447, $p$ < .001), and ingroup-directed comments were more likely to be flagged than comments directed at the outgroup (OR = 1.9, $p$ = .012). As model-based odds ratios can be inflated when comparison base rates are very low, as is the case here given the extremely low flagging rates of critical speech, we also computed raw odds ratios from observed proportions. These showed that the odds of flagging abuse were 69.5 times the odds of flagging criticism, and that the odds of flagging ingroup-directed abuse were 1.31 times the odds of flagging outgroup-directed abuse.

Main effects were qualified in a three-way interaction ($p$ = .003). Estimated marginal means confirmed that Democrats flagged criticism and abuse to a similar extent for ingroup and outgroup targets ($p$ = .272 for criticism and .879 for abuse), while Republicans made more flags for both criticism and abuse directed at the ingroup compared to the outgroup ($p < .001$ for criticism and .002 for abuse). Therefore, the ANOVA and the GLMM produced the same pattern of results for this study. Full statistical outputs for all studies are provided in Supplementary Information.

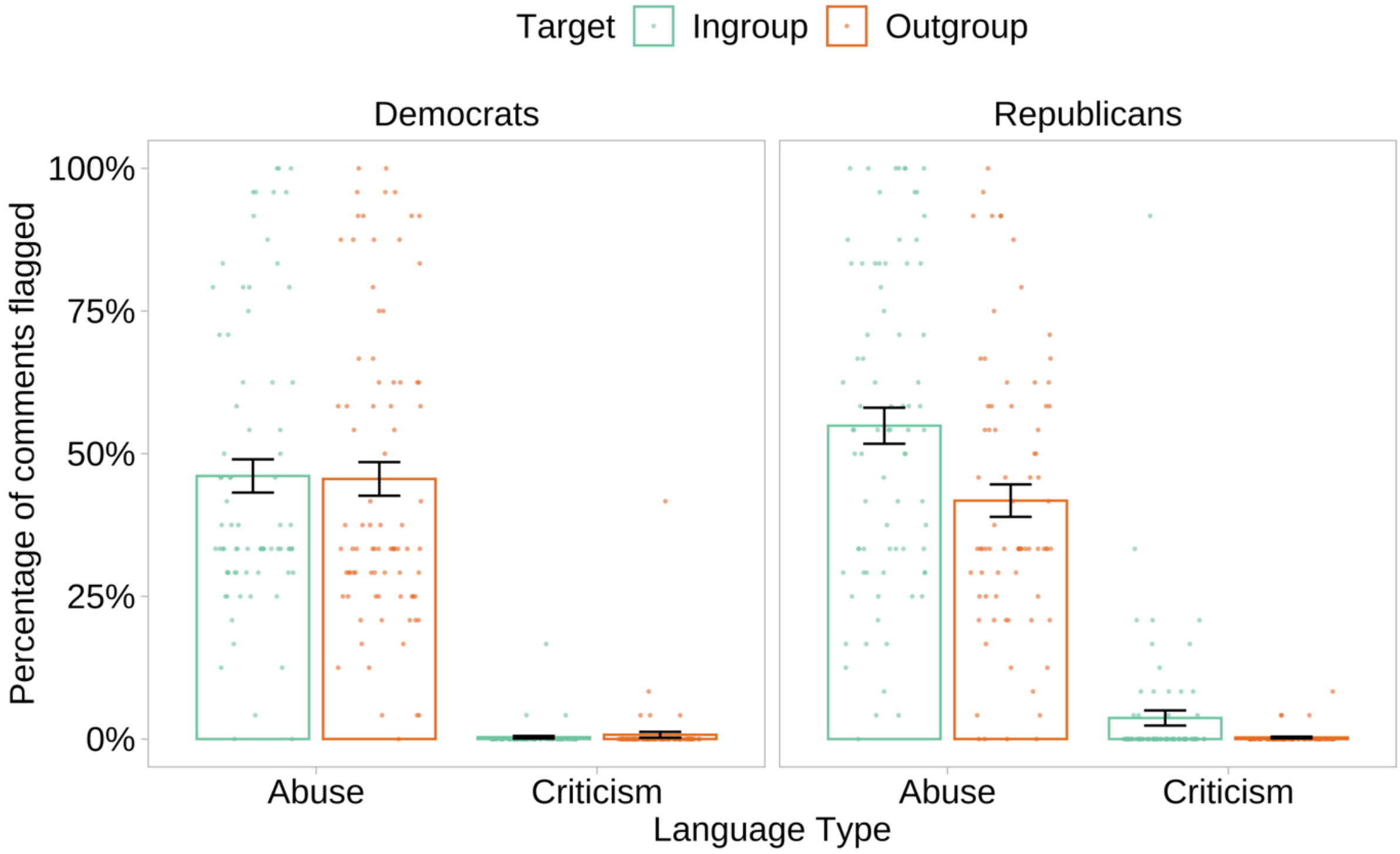

Figure 1: Proportion of flags made for abuse and criticism towards ingroup and outgroup targets by political affiliation. Republican participants flagged abusive and critical comments directed at the ingroup to a greater extent than comments directed at the outgroup, while this ingroup bias effect was not present in Democrats.

**Study 2: Ingroup bias in flagging abuse about vaccination beliefs**

In Study 1, we found that participants flagged abuse directed at the ingroup to a greater extent than abuse directed at the outgroup, but that this ingroup bias was only present amongst Republican participants. We next aimed to examine patterns of flagging behaviours in the context of views about vaccination. We chose this as an additional context known to provoke heated debate online (37). Data was collected in September 2023, during a time in which public debate about vaccines remained salient following the COVID-19 vaccination rollouts (38).

## Methods

### Participants

340 participants completed the study, based on the same power analysis as reported for Experiment 1. Participants were all 18 or over, fluent in English, and based in the US. In line with our pre-registration, we excluded 24 participants that failed at least one of the two attention checks. Of the final sample of 316, 164 participants identified as female, 146 as male, 5 as non-binary and 1 self-described. Participants were aged from 18 to 76 with a mean age of 39.8 (SD = 13.2). 162 participants identified as pro-vaccination, and 154 participants identified as anti-vaccination.

### Task, Procedure and Design

The flagging task, experimental procedure and study design were identical as reported for Experiment 1, other than comments being adapted and directed at people who are pro vaccination or people who are against vaccination. The experimental items were kept identical other than the change in targets (and reference to voters). To set the context for the flagging task, participants this time read a short excerpt from a news article about a clash between pro-vaccine supporters and anti-vaccination protesters at a demonstration in Los Angeles. As for Experiment 1, screening for eligibility was through Prolific, but participants were also asked about their vaccine stance (For or Against) to ensure that their response matched their affiliation recorded on Prolific and that they were in the correct version of the experiment. If they indicated ‘For’ and were in the Against Vaccination version of the study, or vice versa, they were screened out on the basis that the response was inconsistent with their Prolific pre-screening.

**Results and discussion**

A 2(Language type: Abuse/Critical, within subjects) x 2(Target: Ingroup/Outgroup, between subjects) x 2 (Vaccine view: Pro/Anti, between subjects) mixed ANOVA measured differences between proportion of comments flagged in each condition[1].

There was a main effect of language type, $F(1, 312) = 1129.84$, $p < .001$, $\eta_p^2 = 0.784$, showing abuse was flagged to a significantly greater extent than criticism (52.4% of overall abuse was flagged compared to 1.6% of criticism). There was also a main effect of target, $F(1, 312) = 5.49$, $p = .020$, $\eta_p^2 = .017$, showing flagging rates were higher for comments directed at the ingroup than at the outgroup (55.3% of ingroup abuse was flagged compared to 49.4% of outgroup abuse, and 2.8% of ingroup criticism was flagged compared to 0.4% of outgroup criticism).

There was no significant main effect of vaccination view group, $F(1, 312) = 1.69$, $p = .194$, $\eta_p^2 = .005$, showing pro-vaccination and anti-vaccination participants flagged to a similar extent overall. None of the interactions were significant (all *ps* > .05).

Overall, results from the mixed ANOVA show that comments directed at the ingroup were flagged to a greater extent than equivalent comments directed at the outgroup, both for abusive and critical speech, and this was not dependent on group affiliation. However, as for Study 1, the assumptions underlying the ANOVA were not fully met, with residual diagnostics indicating departures from normality and a violation of homogeneity of variance for the critical language condition. We again complemented the ANOVA with a mixed-effects logistic regression fitted to the trial-level binary responses, set up in the same way as reported for Study 1.

[1] We note here that the primary analysis pre-registered was the mixed effects logistic regression and the ANOVA reported was pre-registered as an alternative analysis. For consistency in reporting across studies, we report first results from the ANOVA and then results from the GLMM.

**Mixed-Effects Logistic Regression**

Results from the GLMM extended our key findings. Abusive language was significantly more likely to be flagged than critical language (OR = 616, $p < .001$), and comments directed at the ingroup were significantly more likely to be flagged than those directed at the outgroup (OR = 2.73, $p < .001$). Raw odds ratios calculated from observed proportions showed that the odds of flagging abuse were 67.3 times the odds of flagging criticism and the odds of flagging ingroup-directed abuse were 1.27 times the odds of flagging outgroup-directed abuse.

This time, we observed a significant three-way interaction between language type, target, and vaccination view qualified the main effects ($p = .036$). Estimated marginal means showed that pro-vaccination individuals were no more likely to flag abuse directed at the ingroup than the outgroup ($p = .30$), but were more likely to flag ingroup-directed criticism ($p < .001$). Anti-vaccination individuals were significantly more likely to flag abuse directed at the ingroup ($p =.05$) and marginally more likely to flag criticism directed at the ingroup than the outgroup ($p = .08$).

Taken together, the ANOVA identified only main effects of language type and target, with no evidence that ingroup bias differed between vaccination view groups. In contrast, the GLMM revealed a significant three-way interaction, with follow-up comparisons indicating that the nature of the ingroup bias differed across groups:

pro-vaccination participants showed ingroup bias for critical but not abusive language, whereas anti-vaccination participants showed ingroup bias for abusive language but only marginally for critical language. Because the GLMM is better suited to the binary nature of the trial-level data, we place greater weight on these results.

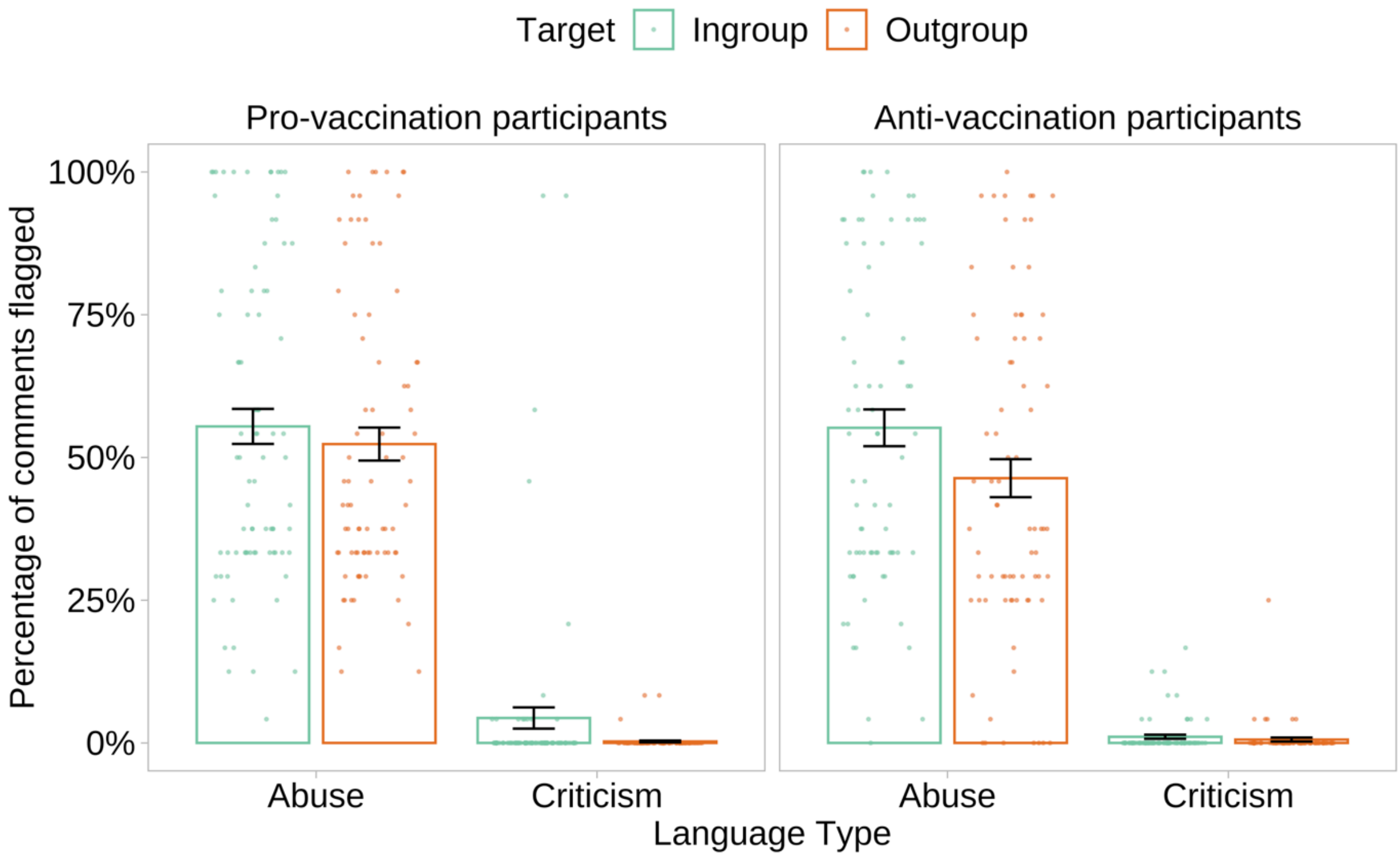

Figure 2: Proportion of flags made for abuse and criticism towards ingroup and outgroup targets by vaccination view affiliation. Follow-up comparisons from the GLMM showed that pro-vaccination participants were significantly more likely to flag criticism directed at the ingroup than the outgroup (but not abuse), while anti-vaccination participants were significantly more likely to flag abuse directed at the ingroup than the outgroup (but only marginally more likely to flag criticism).

## Study 3: Ingroup bias in flagging attacks on beliefs about climate change

In Study 2, pro-vaccination and anti-vaccination participants flagged comments directed at the ingroup more than those directed at the outgroup. We next examine patterns of flagging behaviours in a further context often at the centre of online animosity – views about human-made climate change. Data collection took place in October 2023, when debates about human-made climate change and associated misinformation remained highly salient online (39).

### Methods

### Participants

340 participants completed the study, based on the same power analysis as reported for Experiment 1. Participants were all 18 or over, fluent in English, and based in the US. In line with our pre-registration, we excluded 20 participants that failed at least one of the two attention checks. Of the final sample of 320, 160 participants identified as female, 150 as male, 8 as non-binary and 2 self-described. Participants were aged from 19 to 76 with a mean age of 41.8 (SD = 13.7). 160 participants identified as believing in human-made climate change (believers), and 160 participants identified as not believing in human-made climate change (non-believers).

## Task, Procedure and Design

The flagging task, experimental procedure and study design were identical as reported for Experiment 1, other than comments being directed at climate change activists or people who do not believe in climate change. The experimental items were kept identical other than the change in targets. To set the context for the flagging task, participants this time read a short excerpt from a news article about Climate Week in New York. As before, screening for eligibility was through Prolific, but participants were also asked about their views on human-made climate change (belief in human made climate change, Yes or No) to ensure that their response matched their affiliation recorded on Prolific and that they were in the correct version of the experiment. If they indicated 'No' and were in the Climate Change Believer version of the study, or vice versa, they were screened out on the basis that the response was inconsistent with their Prolific pre-screening.

## Results and discussion

A 2(Language type: Abuse/Critical, within subjects) x 2(Target: Ingroup/Outgroup, between subjects) x 2 (Climate belief: Believer/Non-believer, between subjects) mixed ANOVA measured differences between proportion of comments flagged in each condition.

There was a main effect of language type, $F(1, 316) = 1135.77$, $p <.001$, $\eta_p^2 = 0.782$, showing abuse was flagged to a significantly greater extent than criticism (52.5% of overall abuse was flagged compared to 2.5% of criticism). There was also a main effect of target group, $F(1, 316) = 10.62$, $p =.001$, $\eta_p^2 = .033$, showing flagging rates were higher for comments directed at the ingroup than at the outgroup (56.5% of ingroup abuse was flagged compared to 48.6% of outgroup abuse, and 4.6% of ingroup criticism was flagged compared to 0.5% of outgroup criticism). There was also a significant main effect of climate belief affiliation, $F(1, 316) = 8.47$, $p = .004$, $\eta_p^2 = .026$, showing those who believe in climate change flagged to a greater extent than those who do not (believers flagged 30.2% of all comments, deniers flagged 24.9% of all comments). There was a significant interaction between climate view and language type, $F(1, 316) = 17.74$, $p < .001$, $\eta_p^2 = .053$, which was not relevant to our key research questions and showed that while criticism was flagged to a similar extent by both groups, climate change believers flagged abuse to a greater extent than non-believers. None of the other interactions were significant (all *ps* > .05).

Overall, comments directed at the ingroup were flagged to a greater extent than equivalent comments directed at the outgroup, both for abusive and critical language, and this was not dependent on group affiliation. However, the assumptions underlying the ANOVA were again not fully met, with residual diagnostics indicating departures from normality and a violation of homogeneity of variance for the critical language condition. As before, we complemented the ANOVA with a mixed-effects logistic regression fitted to the trial-level binary responses, set up in the same way as reported above.

**Mixed-Effects Logistic Regression**

Results from the GLMM supported our key findings: Abusive language was significantly more likely to be flagged than critical language (OR = 436, $p < .001$), and comments directed at the ingroup were significantly more likely to be flagged than comments directed at the outgroup (OR = 3.25, $p < .001$). There was also a significant effect of climate view, showing climate change believers made more flags overall than climate change deniers (OR = 2.04, $p = .006$). Raw odds ratios showed that the odds of flagging abuse were 42.4 times the odds of flagging criticism and the odds of flagging ingroup-directed abuse were 1.37 times the odds of flagging outgroup-directed abuse. Additionally, the odds of flagging content for climate change were 1.31 times the odds of flagging content for climate change deniers.

There was a significant interaction between language type and target ($p$=.003). Estimated marginal means showed that both abuse ($p = .002$) and criticism ($p$ <.001) were flagged to a greater extent when directed at the ingroup than the outgroup, but this ingroup bias effect was overall greater for criticism (OR = 1.90 for abuse and 5.56 for criticism). None of the other interactions were significant (all $ps$ > .05). Therefore, the ANOVA and the GLMM produced the same substantive pattern of results for this study: ingroup-directed comments were more likely to be flagged than outgroup-directed comments, and there was no evidence that ingroup bias differed according to climate belief.

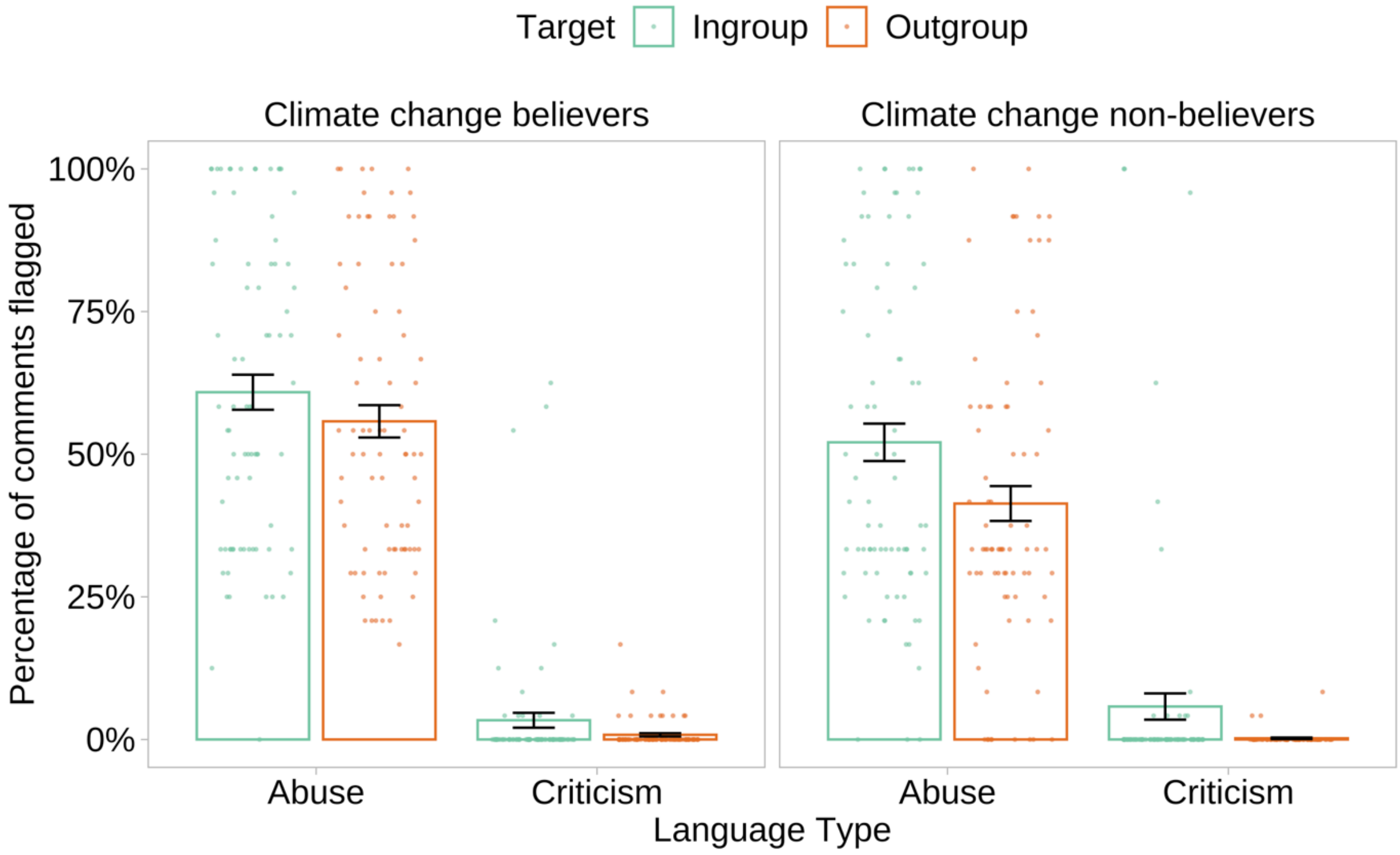

Figure 3: Proportion of flags made for abuse and criticism towards ingroup and outgroup targets by climate change belief affiliation. Comments directed at the ingroup were flagged to a greater extent than equivalent comments directed at the outgroup, both for abusive and critical language.

### Study 4a: Ingroup bias in flagging misogynistic hate speech

Studies 1-3 examine user reports of abuse in three distinct contexts, political affiliation, vaccination opinions, and climate change beliefs. In Study 4a and 4b, we examine patterns of flagging behaviours when language is reminiscent of a more severe and widespread type of online abuse – misogynistic hate (40). We present examples of misogynistic hate in the context of the widely-debated topic of abortion rights in the US (41). Data collection for Experiments 4a and 4b took place in March 2023. This was at a time when abortion rights remained a highly contentious public and political issue in the U.S. following the over-turning of *Roe v. Wade* the previous year and during the resulting wave of state-level restrictions.

## Methods

### Participants

340 participants completed the study, based on the same power analysis as reported for Study 1. Participants were all 18 or over, fluent in English, and based in the US. In line with our pre-registration, we excluded 21 participants that failed at least one of the two attention checks. Of the final sample of 319, 153 participants identified as female, 154 as male, 11 as non-binary and 1 self-described. Participants were aged from 18 to 93 with a mean age of 40.0 (SD = 14.5). 158 participants identified as pro-abortion rights, and 161 participants identified as anti-abortion.

### Task, Procedure and Design

The flagging task, experimental procedure and study design were similar as reported for Experiment 1, other than comments being directed at women who are pro abortion rights (sometimes referred to as 'pro-choice') or women who are anti-abortion (sometimes referred to as 'pro-life'). The experimental items were similar to those presented in Experiments 1-3, however some of the language was altered to reflect typical examples of misogynistic hate online (e.g., including threats of sexual assault, comparisons to cows and dogs, and references to promiscuity or hysteria) and to reflect the specific context of the comments (e.g., caring about the life of children/the life of women). All experimental items are available on the OSF page for this work. To set the context for the flagging task, participants this time read a short excerpt from a news article about a clash between pro- and anti-abortion groups at a planned parenthood clinic in California. Screening for eligibility was again through Prolific and participants were also asked about their views on abortion rights (For or Against) to ensure that their response matched their view recorded on Prolific. If they indicated ‘Against’ and were in the For Abortion Rights version of the study, or vice versa, they were screened out on the basis that the response was inconsistent with their Prolific pre-screening.

**Results and discussion**

A 2(Language type: Abuse/Critical, within subjects) x 2(Target: Ingroup/Outgroup, between subjects) x 2 (Abortion view: Pro-abortion rights / Anti-abortion, between subjects) mixed ANOVA measured differences between proportion of comments flagged in each condition.

There was a main effect of language type, $F(1, 315) = 1405.69$, $p < .001$, $\eta_p^2 = .817$, showing abuse was flagged to a significantly greater extent than criticism (59.6% of overall abuse was flagged compared to 3.0% of criticism). There was also a main effect of target, $F(1, 315) = 20.21$, $p < .001$, $\eta_p^2 = .06$, showing flagging rates were higher for comments directed at the ingroup than at the outgroup (65.4% of ingroup abuse was flagged compared to 53.7% of outgroup abuse, and 5% of ingroup criticism was flagged compared to 0.98% of outgroup criticism). There was no significant main effect of group affiliation, $F(1, 315) = 1.81$, $p = .180$, $\eta_p^2 = .006$, with pro- and anti-abortion rights individuals flagging to a similar extent.

There was a significant interaction between language type and abortion view, $F(1, 315) = 8.32$, $p = .004$, $\eta_p^2 = .026$ and between language type and target, $F(1, 315) = 6.73$, $p = .010$, $\eta_p^2 = .021$, though these effects were not relevant for our key research questions[2]. None of the other interactions were significant (*ps* > .05).

Overall, comments directed at the ingroup were flagged to a greater extent than equivalent comments directed at the outgroup, and this was not dependent on group affiliation. However, the assumptions underlying the ANOVA were again not fully met, with residual diagnostics indicating departures from normality and a violation of homogeneity of variance for both language conditions. As before, we complemented the ANOVA with a mixed-effects logistic regression.

**Mixed-Effects Logistic Regression**

The GLMM, set up in the same way as reported for Studies 1-3, confirmed our key findings. Abusive language was significantly more likely to be flagged than critical language (OR = 259, $p < .001$), and comments directed at the ingroup were significantly more likely to be flagged than comments directed at the outgroup (OR = 2.66, $p < .001$). There was no main effect of group affiliation ($p = .633$). Raw odds ratios showed that the odds of flagging abuse were 47.9 times higher than for criticism, and the odds of flagging ingroup-directed abuse were 1.63 times higher than for outgroup-directed abuse.

---

[2] Pairwise comparisons on the language type*abortion view interaction showed that pro-abortion rights individuals flagged abuse to a greater extent than anti-abortion rights participants overall, while there was no difference in the extent to which each group flagged criticism. For the language type*target interaction, pairwise comparisons show that while ingroup-directed comments were flagged to a significantly greater degree than outgroup-directed comments for both abuse and criticism, this bias effect was stronger for abuse.

There was an interaction between language type and abortion view ($p < .001$) and the three-way interaction between language type, target group and group affiliation was marginally significant ($p = .055$). Post hoc comparisons of estimated marginal means showed that pro-abortion participants flagged hate directed at the ingroup more than hate directed at the outgroup ($p < .001$) but this effect was only marginal for critical speech ($p = .058$). Anti-abortion rights participants showed a significant ingroup bias for both hate and critical speech ($p = .011$ for hate; $p < .001$ for criticism). Taken together, the ANOVA and GLMM yielded the same substantive conclusions. Both analyses indicated that ingroup-directed comments were more likely to be flagged than outgroup-directed comments, with little evidence that this ingroup bias depended on abortion view. Although the GLMM identified a marginal three-way interaction, this did not alter the overall interpretation of the results.

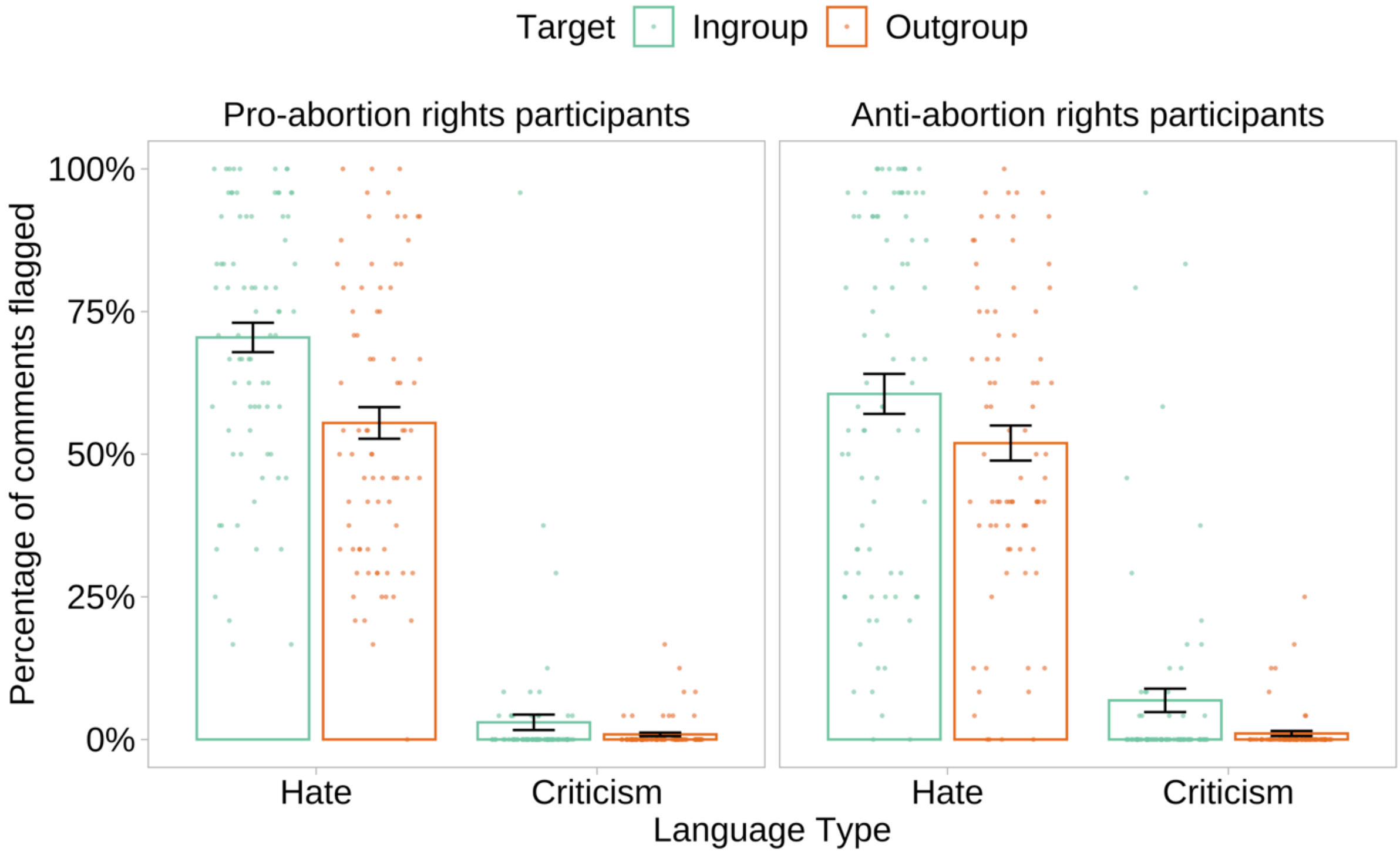

Figure 4: Proportion of flags made for hate and criticism towards ingroup and outgroup targets by abortion stance affiliation. Hate and criticism directed at the ingroup were flagged to a greater extent than equivalent comments directed at the outgroup.

### Study 4b: Ingroup bias in flagging misogynistic hate speech, within subjects replication

In Studies 1-4a, participants saw abuse and criticism directed at either ingroup or outgroup targets. While much online conversation may be polarised in this way (42), it is also likely that there are times when people are exposed to hate and abuse directed at multiple targets on opposing sides of online debate. In Study 4b, we examine patterns of flagging when participants are exposed to hate and criticism against both ingroup and outgroup targets.

## Methods

### Participants

340 participants completed the study, based on the power analysis as reported for Experiment 1. Participants were all 18 or over, fluent in English, and based in the US. In line with our pre-registration, we excluded 30 participants that failed at least one of the two attention checks. Of the final sample of 310, 155 participants identified as female, 148 as male, 4 as non-binary, 2 self-described and 2 chose 'prefer not to say'. Participants were aged from 19 to 84 with a mean age of 41.8 (SD = 14.9). 156 participants identified as pro-abortion rights, and 154 participants identified as anti-abortion.

### Task, Procedure and Design

The flagging task, experimental procedure and study design were similar as for Experiment 4a but with one key exception: this time, instead of seeing comments directed at either pro- or anti-abortion rights women, participants saw abuse and criticism directed at both targets in the same series of comments under the news story.

While different comments were directed at each group, the level of abuse and criticism was matched exactly for the two targets. To do this, we took the same 48 experimental items (24 abuse and 24 criticism) from Study 4a, and we divided these in two such that we had two distinct sets of sentences, each with 12 examples of abuse and 12 examples of criticism. We counterbalanced participants such that half of the participants saw sentence set 1 directed at pro-abortion rights women and sentence set 2 directed at anti-abortion women, and half of the participants saw sentence set 1 directed at anti-abortion women and sentence set 2 directed at pro-abortion rights women. This meant that while individual participants saw different comments for each target, across the experiment, the comments towards both targets were matched. The 48 experimental items (hate and criticism for both targets) were presented together in a randomised order, along with five filler items and two attention checks, as before. This results in a 2 (Language type: abuse/criticism, within-subjects)*2 (Target: ingroup/outgroup, within-subjects)*2 (Political affiliation: Pro-abortion rights/Anti-abortion between-subjects) mixed factorial design.

**Results and discussion**

As pre-registered, a 2(Language type: Abuse/Critical, within subjects) x 2(Target: Ingroup/Outgroup, within subjects) x 2 (Abortion view: Pro-abortion rights / Anti-abortion, between subjects) mixed ANOVA measured differences between proportion of comments flagged in each condition.

In line with our experimental manipulation, there was a main effect of language type, $F(1, 308) = 1163.89$, $p < .001$, $\eta_p^2 = .791$, showing abuse was flagged to a significantly greater extent than criticism (58.5% of overall abuse was flagged compared to 3.5% of criticism). There was also a main effect of target, $F(1, 308) = 17.55$, $p < .001$, $\eta_p^2 = .054$, showing flagging rates were higher for comments directed at the ingroup than at the outgroup (60.4% of ingroup abuse was flagged compared to 56.6% of outgroup abuse, and 4.5% of ingroup criticism was flagged compared to 2.5% of outgroup criticism).

There was no significant main effect of group affiliation, $F(1, 308) = 0.01$, $p = .909$, $\eta_p^2 = < .001$, showing that pro-abortion rights and anti-abortion individuals flagged to a similar extent. While none of the two-way interaction terms were significant (all *ps* $> .05$), there was a significant three-way interaction between language type, abortion view and target, $F(1, 308) = 16.12$, $p < .001$, $\eta_p^2 = < .05$. Pairwise comparisons showed that pro-abortion rights participants flagged ingroup-directed abuse to a greater extent than outgroup-directed abuse, ($M_{IG} = 63.1\%$, $M_{OG} = 56.8\%$, $p < .001$), but flagged critical comments to a similar extent for both target groups ($M_{IG} = 2.1\%$, $M_{OG} = 1.6\%$, $p = .325$). Anti-abortion rights participants flagged abuse directed at both target groups to a similar extent ($M_{IG} = 57.7\%$, $M_{OG} = 56.4\%$, $p = .419$), but flagged criticism directed towards the ingroup to a greater extent than criticism directed towards the outgroup ($M_{IG} = 6.9\%$, $M_{OG} = 3.4\%$, $p = .004$). Overall, comments directed at the ingroup were flagged to a greater extent than equivalent comments directed at the outgroup, but for pro-abortion rights participants this ingroup bias effect was only seen in abusive comments, and for anti-abortion rights participants this effect was only seen in critical comments. As reported for Studies 1-4a, the ANOVA assumptions were not fully met, with residual diagnostics indicating departures from normality and a violation of homogeneity of variance for the ingroup-directed criticism condition only. As before, we complemented the ANOVA with a mixed-effects logistic regression.

**Mixed-Effects Logistic Regression**

In our final mixed-effects logistic regression, we again included fixed effects of language type (hate vs. critical), target (ingroup vs. outgroup), and abortion view (pro-choice vs. pro-life), along with their interactions, and the categorical predictors were sum coded as for previous models.

The model revealed that abusive language was significantly more likely to be flagged than critical language (OR = 162, $p <.001$), and comments directed at the ingroup were significantly more likely to be flagged than comments directed at the outgroup (OR = 1.51, $p < .001$). There was also a main effect of abortion view such that anti-abortion participants were more likely to flag comments than pro-abortion rights participants (OR = 1.62, $p=.018$). Raw odds ratios showed that the odds of flagging abuse were 39.3 times the odds of flagging criticism, while the odds of flagging ingroup-directed abuse were 1.17 times the odds of flagging outgroup-directed abuse. The odds of flagging among anti-abortion participants were 1.01 times the odds among pro-abortion rights participants, indicating a negligible difference in overall flagging behaviour, consistent with the non-significant main effect observed in the ANOVA.

These main effects were qualified in a significant three-way interaction ($p$ = .007). Post hoc comparisons of estimated marginal means, in line with our initial ANOVA result, showed that pro-abortion rights participants flagged hate to a greater extent when directed at the ingroup than the outgroup ($p$ <.001) but there was no intergroup difference in rates of flagging for critical speech ($p$ =.254). Anti-abortion rights participants flagged criticism to a greater extent when directed at the ingroup than the outgroup ($p$ <.001) but there was no intergroup difference in rates of flagging for hate ($p$ = .337). As such, the ANOVA and the GLMM produced the same pattern of results, with both showing that pro-abortion rights participants exhibited ingroup bias for abusive language, while anti-abortion participants exhibited ingroup bias for critical language. Although the GLMM additionally identified a significant main effect of abortion view, this did not alter the overall interpretation of the findings.

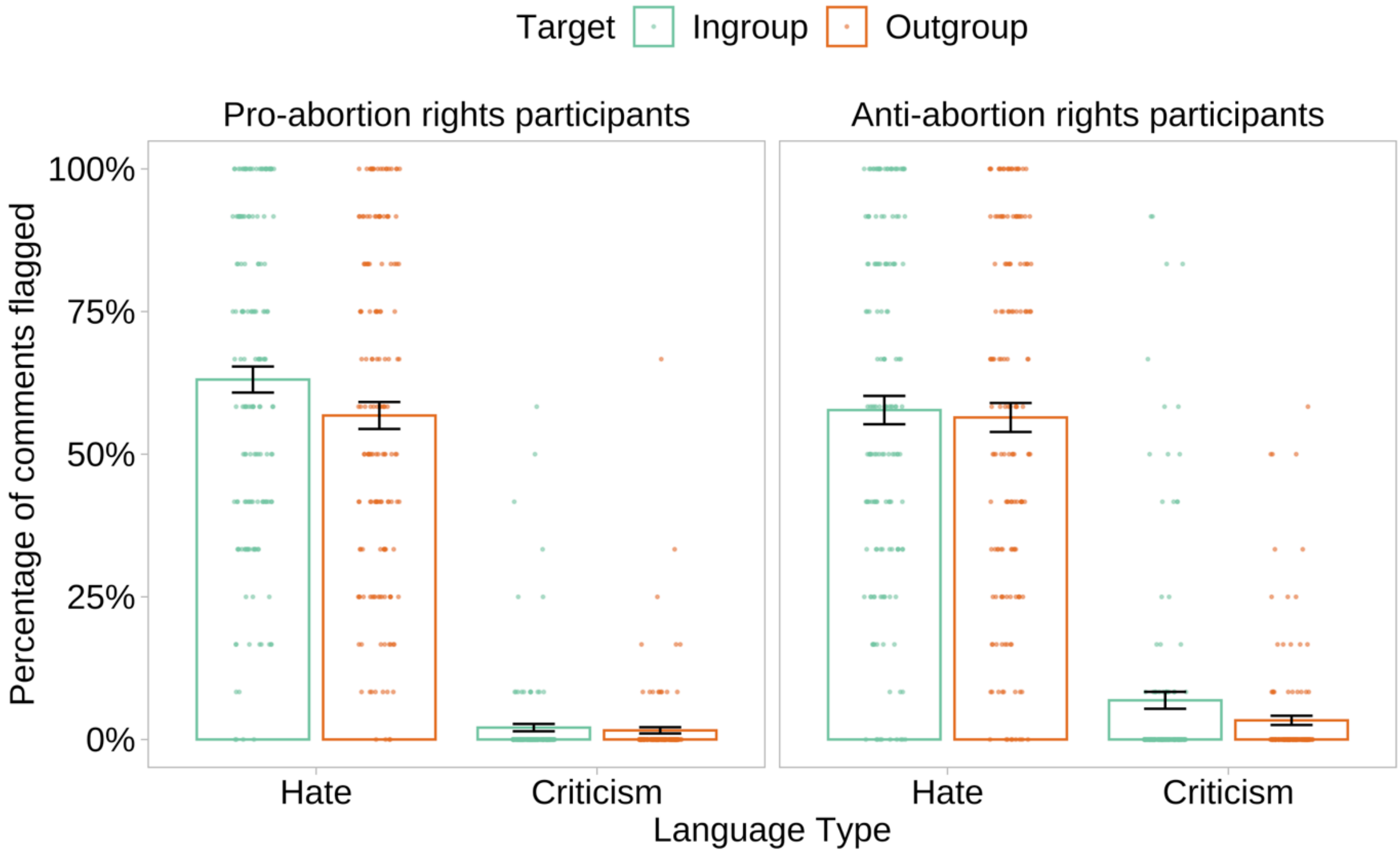

Figure 5: Proportion of flags made for hate and criticism towards ingroup and outgroup targets by abortion stance affiliation in a within-subjects design. Comments directed at the ingroup were flagged to a greater extent than comments directed at the outgroup, but for pro-abortion rights participants this ingroup bias effect was only seen in hateful comments, and for anti-abortion rights participants this effect was only seen in critical comments.

## General discussion

Across five pre-registered and well-powered experiments, we examined people's engagement with reporting hate and abuse online in four distinct intergroup contexts: political affiliation, views on vaccination, beliefs about climate change, and the abortion rights debate. We chose these contexts to reflect a range of topics and contexts which provoke heated debate and polarisation in online discussion. Across all experiments, participants typically reported 50-60% of abusive comments overall, and less than 5% of criticism. Little is currently known about the extent to which people flag hate and abuse in real online environments, though survey work in the UK suggests that while a majority (about 60%) of social media users who are aware they can do so have flagged content to a platform at some point (24), far fewer do so regularly and in response to all harms they encounter (25). In our studies, which included clear instructions and reminders about using the flagging function, we observed reliable flagging of abusive content and low rates of reporting criticism. This pattern has potential implications for improving flagging rates online, an issue of direct interest for policymakers (43). Simple design changes such as clear guidance and active encouragement may significantly boost engagement with flagging tools and help platforms identify harmful content more efficiently. While our findings were observed in a controlled experimental setting in which participants were financially incentivised to complete the task, they provide initial indication that guidance and encouragement could play a role in enhancing uptake of flagging tools. Future work should explore how such interventions operate in real-world social media environments.

In all four social contexts that we examined, we found a robust ingroup bias effect in flagging, where comments directed at the ingroup were flagged to a greater extent than comments directed at the outgroup. As well as flagging outgroup-directed abuse to a lesser extent than ingroup-directed abuse, participants were also more likely to misclassify criticism directed at the ingroup as content that should be flagged. These findings align with previous work showing political biases are common in crowdsourced flagging of misinformation (26). They also complement evidence from offline settings showing that people's endorsement of police intervention in the context of a demonstration depended on whether the demonstration was perceived to ideologically in line with participants' own views or not (28). Our work is in line with a body of findings in social cognition which show that social identity shapes multiple levels of cognition, from low level perception to higher level attitudes and behaviours (29,30,32,33) and provides novel evidence that ingroup bias extends to user engagement with reporting hate and abuse online. Our results show that when designing interventions aimed at improving flagging rates, efforts should focus not only on increasing overall engagement with these tools, but also on enhancing the accuracy and fairness of reports.

However, although abuse towards the outgroup was flagged less, it was still flagged relatively consistently overall (for example, in Study 1, 50.5% of ingroup abuse was flagged compared to 43.8% of outgroup abuse and in the context of language containing misogynistic hate in Study 4a, 65.4% of ingroup abuse was flagged compared to 53.7% of outgroup abuse). This pattern is promising in that despite exhibiting a persistent ingroup bias, participants did recognise a substantial portion of outgroup-directed hate and abuse as content that should be flagged. This suggests that even in highly polarised intergroup contexts, users are able to make accurate judgments about harmful language. The ingroup bias did not undermine participants' abilities to identify and report hate and abuse overall, rather, the ingroup bias subtly shifts sensitivity to reporting abuse alongside robust differentiation between abuse and criticism.

In our final experiment, when participants were presented with comments directed at both ingroup and outgroup members, ingroup bias effects were weaker but still present. This suggests that implicitly reminding participants of the shared experience of abuse across groups (by exposing them to all comment types in the same feed), narrowed the bias effect. Specifically, flagging for ingroup-directed hate was reduced, and flagging for outgroup-directed hate was increased (in relation to Study 4a). This pattern raises the possibility that exposure to harmful content directed at multiple targets may reduce the influence of ingroup bias. Practically, professional content moderators or trained flaggers who routinely encounter hate and abuse directed at multiple groups may be less susceptible to ingroup biases than typical users operating in potentially polarised environments. Future work should directly test how different moderation environments shape the emergence and attenuation of ingroup biases in practice.

While participants consistently flagged abusive comments to a greater extent than critical ones and comments directed at the ingroup to a greater extent than comments directed at the outgroup, patterns were subtly different across the four social contexts. In the context of politically motivated abuse in the US, the ingroup bias in flagging was only present amongst Republican participants, while Democrats flagged ingroup-directed and outgroup-directed abuse to a similar extent. In the contexts of opinions on vaccination, beliefs about climate change, and abortion rights, all participants exhibited an ingroup bias in flagging. Climate change believers flagged abusive content more frequently than climate change deniers, and notably, in the abortion rights context, where misogynistic hate was presented, overall flagging rates for hate were higher than in the other contexts, indicating heightened engagement when language is particularly emotive. While some patterns of flagging differed subtly across group affiliations, it is difficult to draw firm conclusions about the sources of these asymmetries without further evidence. Some work suggests that right-leaning individuals may be less willing to engage with moderation tools (44). However, we do not include measures of political orientation and so cannot test this explanation. Notably, although group positions are often associated with political ideology, we did not observe a consistent typical left–right pattern across studies. We cannot rule out the possibility that the news articles, while selected to be as neutral as possible, were implicitly more favourable to some groups than others, or that the timing of data collection coincided with salient events affecting particular groups. Additionally, differences in the strength of group identification across contexts may have shaped responses, which we discuss further below.

Our work makes a novel contribution by providing the first systematic experimental evidence that ingroup bias shapes user engagement with reporting hate and abuse online. Our findings offer valuable insights for policymakers and platform designers aiming to develop interventions that enhance not only the frequency but also the fairness and accuracy of user-generated flags. However, several limitations should be noted. Results are obtained from a basic mock social media feed implemented in Qualtrics which does not reproduce all the features of a typical social media feed, for example usernames and profile pictures, number of shares and indications of 'Likes' and other reactions. Future work would benefit from building on this approach by embedding the task in more interactive mock-platform environments to further enhance ecological validity. It is also worth noting that any results obtained in a controlled experimental setting may differ from real-world online behaviours where many interacting factors are likely to be at play, such as distraction, a more rapid pace of content consumption, peer dynamics and individual platform design features. It would also be interesting for future work to examine the potential role of desensitization in responses to online hate. While some work suggests that frequent encounters with hateful or abusive content can reduce emotional reactivity, lowering bystander intervention or reporting behaviour (45,46), other studies suggest that the opposite may occur, with exposure to hate heightening negative emotional responses over time (47). Clarifying the direction of these effects and how they interact with user reporting is important for informing interventions designed to protect both users and bystanders from online hate and abuse.

We focus exclusively on text-based hate and abuse, while images and videos are increasingly common on social media. We also did not analyse or control for the potential moderating effects of demographic variables such as age, gender, education level or digital literacy on overall flagging or on biases within flagging behaviours, which may be of interest for future work. For example, recent work finds that women are more likely to use some forms of safety technology than men, perhaps because of heightened fears for safety online (6,24). Work elsewhere shows that while older adults tend to be more prosocial in general, they also tend to display heightened favouritism for ingroup members (48). It would be interesting to explore the potential interacting effects of demographic characteristics on flagging behaviours more broadly.

Participants were classified into group affiliations for each experiment using single self-report screening items but we did not measure their strength of identification with the groups. While our categorisation approach is consistent with prior work measuring various effects of group identity (28,49–52), it is plausible that strength of identification may moderate reactions to ingroup and outgroup-directed online hate and abuse. For example, strength of group identification has been shown to motivate certain forms of pro-ingroup action (53). Future work would benefit from testing whether stronger or more fused social identities further amplify flagging responses to ingroup-directed hate and abuse. Finally, our sample was restricted to the U.S., limiting the generalisability of results to other cultural contexts, and our samples were not designed to be nationally representative, meaning results cannot be taken as estimates of population-level flagging rates. Future research should address these gaps by examining diverse media formats, broader populations, and real-world flagging behaviours to develop a deeper understanding of improving user engagement with these tools.

Taken together, our studies suggest that when primed to engage with online flagging mechanisms, people flag abuse reliably across a variety of online contexts. Additionally, while people consistently flag ingroup-directed abuse to a greater extent than outgroup-directed abuse, this effect is weakened when people are implicitly reminded of the range of targets than may suffer similar levels of abuse. Importantly, there are also subtle differences in flagging behaviours across the type of online discussion (political, scientific, and moral) and the group affiliation of the user. Our findings provide a novel perspective on the nuanced ways in which people engage with flagging hate and abuse online and have important implications for the design of effective interventions to improve flagging behaviours.